\documentclass[12pt, preprint]{aastex}
\begin{document}
\title{DETECTION OF THE VELOCITY SHEAR EFFECT ON THE SPATIAL DISTRIBUTIONS OF THE GALACTIC 
SATELLITES IN ISOLATED SYSTEMS}
\author{Jounghun Lee\altaffilmark{1}, Yun-Young Choi\altaffilmark{2}}
\altaffiltext{1}{Astronomy Program, Department of Physics and Astronomy, 
Seoul National University, Seoul 151-747, Korea; jounghun@astro.snu.ac.kr}
\altaffiltext{2}{Department of Astronomy and Space Science, Kyung Hee 
University, Gyeonggi 446-701, Korea; yy.choi@khu.ac.kr}
%%%%%%%%%%%%%%%%%%%%%%%%%%%%%%%%%%%%%%%%%%%%%%%%%%%%%%%%%%%
\begin{abstract}
We report a detection of the effect of the large-scale velocity shear on the spatial distributions of the galactic satellites around 
the isolated hosts. Identifying the isolated galactic systems each of which consists of a single host galaxy and its satellites 
from the Seventh Data Release of the Sloan Digital Sky Survey and reconstructing linearly the velocity shear field in the local 
universe, we measure the alignments between the relative positions of the satellites from their isolated hosts and the principal axes 
of the local velocity shear tensors projected on to the plane of sky. We find a clear signal that the galactic satellites in isolated 
systems are located preferentially along the directions of the minor principal axes of the large-scale velocity shear field.  
Those galactic satellites which are spirals, brighter, located at distances larger than the projected virial radii of the hosts and 
belonging to the spiral hosts yield stronger alignment signals, which implies that the alignment strength depends 
on the formation and accretion epochs of the galactic satellites. 
It is also shown that the alignment strength is quite insensitive to the cosmic web environment as well as the size and luminosity of the 
isolated hosts. Although this result is consistent with the numerical finding of Libeskind et al. based on a $N$-body experiment, 
due to the very low significance of the observed signals it remains inconclusive whether or not the velocity shear effect on the satellite 
distribution is truly universal. 
\end{abstract}
%%%%%%%%%%%%%%%%%%%%%%%%%%%%%%%%%%%%%%%%%%%%%%%%%%%%%%%%%%%
\section{INTRODUCTION}
\label{sec:intro}

The filamentary distributions of the galaxies and the web-like interconnections between the structures over 
different scales are collectively called the cosmic web phenomenon. The recent observations have disclosed the 
ubiquitous appearance of the cosmic web phenomena \citep[e.g.,][]{6dfgrs,sdssdr7,2mrs},  
which agrees with the numerical results from  $N$-body simulations for a standard $\Lambda$CDM cosmology where the 
energy density of the universe is contributed dominantly by the cosmological constant  $\Lambda$ and the collisionless Cold 
Dark Matter \citep[e.g.,][]{web96,springel-etal05}.  
An urgently requested task in the field of the large scale structure is to find a tracer of the cosmic web with which 
the complicated filamentary pattern in the cosmic web can be coherently described and the key mechanism for the 
web interconnection can be quantitatively understood \citep{cautun-etal14,tempel-etal14b}. 

The velocity shear field proposed by \citet{hoffman-etal12} as a new diagnostic of the web environment has 
recently drawn exploring attention. They suggested that the cosmic environments be classified into the four 
standard categories (i.e., knot, filament, sheet, void) by putting some positive threshold on the three eigenvalues of 
the local velocity shear tensor and demonstrated with a high-resolution $N$-body simulation that the web environments 
classified by their algorithm reproduced the expected images of the four categories. 
But, there is no objective way to classify the cosmic web and thus it is quite subjective to claim that  
the velocity shear field is an optimal tracer of the cosmic web.

It is worth mentioning here how the velocity shear field is different from the tidal shear field that was widely used 
as a tracer of the cosmic web in the previous approaches based on the classical cosmic web theory 
\citep[e.g.,][]{web96,hahn-etal07,platen-etal08,lee-etal09,zhang-etal09,zhang-etal13}. 
The tidal shear field defined as the second derivative of the gravitational potential is basically a linear quantity, 
obtained under the implicit assumption that the peculiar velocity field is curl free. Whereas, the velocity shear field 
is an extension of the tidal shear field to the nonlinear regime where the velocity field develops a curl mode (i.e., vorticity). 
Given that the cosmic web pattern penetrates deep into the nonlinear scales, the velocity shear field 
should trace better the nonlinear evolution of the cosmic web, being more suitable for the description of 
the interconnection over widely different scales \citep{hoffman-etal12}.

The usefulness of the velocity shear field as a tracer of the cosmic web has been noted by the subsequent numerical studies 
based on high-resolution N-body experiments.
\citet{libeskind-etal13} explained that  the alignments between the shapes and angular momentum of the dark 
matter halos and the large-scale environments must have been mainly caused by the anisotropic gravitational 
collapse and accretion of dark matter along the minor principal axes of the velocity shear field. 
\citet{libeskind-etal14a} demonstrated that the web interconnection should be a consequence of the stability of 
the velocity shear field against the change of the filtering scale.
\citet{tempel-etal14a} showed that the filamentary structures identified in their $N$-body simulations 
are indeed elongated along the minor principal axes of the velocity shear field.
 
\citet{libeskind-etal14b} performed a more systematic investigation of the velocity shear effect on the generation of 
the cosmic web.  Noting that the presence of the cosmic web phenomena on all scales essentially implies that there 
are preferred directions in the occurrence of the accretion and merging events through which the structure formation 
proceeds in the standard $\Lambda$CDM picture, \citet{libeskind-etal14b} have shown by their $N$-body 
experiments that the infall and accretion events of the satellites always occur along the minor principal axes of the 
velocity shear field on all scales, no matter what environment the satellites are located and no matter what mass the 
satellites and hosts have.  

These works have provided theoretical explanations for the strong alignments between the spatial distribution of the satellites 
and the shapes of their hosts, which were shown by many $N$-body or hydrodynamic simulations 
\citep[e.g.,][]{wang-etal05,zentner-etal05,AB06,kang-etal07,libeskind-etal07,deason-etal11,wang-etal14,dong-etal14}. 
As the accretion of dark matter onto a host occurs along the direction of the minor principal axis of the {\it small-scale} 
velocity shear field, the shape of the host becomes elongated along that direction. 
As the infall of the satellites into the host occurs preferentially along the direction of the minor principal axis of the larger 
scale velocity shear field, the spatial distributions of the satellites around the hosts are aligned with that direction. 
Since the velocity shear field on two different scales are strongly cross-correlated \citep{libeskind-etal14a}, 
the shape of a host  becomes aligned with the spatial distribution of its satellites. 

Although the alignments between the host shapes and the satellite locations have been supported by several observational 
studies \citep[e.g.,][]{brainerd05,yang-etal06,azzaro-etal07,bailin-etal08,AB10,AB11}, 
no direct observational evidence was found for the preferential alignments of the satellite infall directions with the minor principal 
axes of the velocity shears nor for the strong cross-correlations between the velocity shear field smoothed on different scales. 
Very recently, \citet{lee-etal14} have found an observational evidence for the alignments between the infall directions 
of the satellite galaxies and the minor principal axes of the local velocity shear tensors.  Analyzing the spatial distribution of 
the satellite galaxies around the Virgo cluster from the Extended Virgo Cluster Catalog (EVCC) compiled by \citet{kim-etal14} 
and determining the principal directions of the velocity shear tensor around the Virgo cluster from the peculiar velocity field 
recently reconstructed by \citet{wang-etal12},  they showed that the satellite galaxies located beyond the virial 
radius of the Virgo cluster tend to reside in the plane perpendicular to the major principal axes of the local velocity shear tensor. 

In their analysis, however, \citet{lee-etal14} projected the positions of the Virgo satellites as well as the principal 
axes of the local velocity shear tensor onto the plane of the sky to eliminate the severe effect of nonlinear redshift space 
distortion expected to be present around the highly overdense region around the Virgo. But, the minor principal axes of 
the local velocity shear  at the Virgo center turned out to be almost perpendicular to the line of sight direction and thus the true 
three dimensional direction was completely lost during the projection process. In consequence, what \citet{lee-etal14} was able  
to find was not directly the alignments of the Virgo satellites with the minor principal axis of the local velocity shear tensor but 
their anti-alignments with the major principal axis that almost lies in the plane of the sky. 
 
In order to find a more direct observational evidence for the alignments between the satellite locations and the minor 
principal axes of the velocity shear field, it is necessary to consider those systems where the non-linear redshift effect is not 
severe. An isolated galactic system composed of a single host galaxy and its satellites should be a good target for this study.  
The nonlinear redshift distortion effect is expected to be less severe around a galaxy than in a 
cluster. Furthermore, for the case of an {\it isolated} galactic system,  its satellites can be readily identified by analyzing  the 
spatial distributions of the galaxies around it. 
We attempt here to find a signal of the alignments between the locations of the galactic satellites around the isolated host 
galaxies and the minor principal axes of the velocity shear tensors measured at the positions of the hosts by analyzing the 
galaxy catalog from the Seventh Data Release of the Sloan Digital Sky Survey (SDSS DR7) \citep{sdssdr7}. 
Unlike in the work of \citet{lee-etal14}, we will have to take a {\it statistical approach} rather than focusing on each galactic 
system since the number of the satellites around an isolated galaxy is much lower than that around a massive cluster like 
the Virgo cluster. 

The organization of this paper is as follows. In section \ref{sec:data}, we describe the observational data used for our analysis. 
In section \ref{sec:analysis}, we present an observational evidence for the alignments between the spatial distributions of the 
galactic satellites in isolated systems and the minor principal axes of the velocity shear field.  
In section \ref{sec:depend} we explain if and how the strength of the satellite-shear alignments depends on the properties 
of the galactic satellites and their hosts and on the web environments as well. 
In section \ref{sec:con} we discuss our results by comparing them with the recent numerical findings and draw a final conclusion. 
Throughout this Paper, we assume a flat $\Lambda$CDM universe whose initial conditions are described by the key 
cosmological parameters set at the values of $\Omega_{m}=0.26,\ \Omega_{\Lambda}=0.74,\ h=0.73,\ \sigma_{8}=0.8$. 

\section{OBSERVATIONAL DATA}
\label{sec:data}

We construct a sample of the isolated galactic systems each of which consists of a single host and at least one 
satellite from the SDSS DR7 spectroscopic datasets \citep{sdssdr7} by following the procedure prescribed in \citet{HP10}. 
Focusing on the local volume in the redshift range of $0< z < 0.08$, we select those galaxies as the hosts which are 
at least one magnitude brighter in the $r$-band than the survey limit, which amounts to putting a threshold $m_{c}=16.77$ as 
an upper limit on the $r$-band magnitudes, $m_{r}$, of the SDSS galaxies. The fainter galaxies with $m_{r}> m_{c}$ are 
excluded from our sample of the host galaxies since the satellite selection should be very incomplete for the case of those 
fainter galaxies (see below for the satellite selection procedure).  
The absolute $r$-band magnitude, $M_{r}$, of each selected galaxy as an isolated host is calculated 
from $m_{r}$ by properly taking into account both of the $K$-correction \citep{blanton-etal05} and the luminosity 
evolution correction \citep{tegmark-etal04}.

For each host with $m_{r}\le m_{c}$, we look for its nearest neighbor galaxy in the redshift range of $-0.05< z < 0.085$ and 
determine the difference between the heliocentric velocities of the host and its nearest neighbor along the line of sight, 
$\Delta v_{\rm n}$, as well as the projected separation distance between them in the projected plane of the sky, $r_{n}$. 
As done in \citet{HP10},  the host galaxies are identified as the isolated ones if their nearest neighbors satisfy two conditions 
of  $r_{n} \ge {\rm max}\{R_{\rm v},\ R_{\rm vn}\}$ and $\Delta v_{\rm n} \ge \Delta v_{c}$, where $R_{\rm v}$ and $R_{\rm vn}$ are 
the projected virial radii of the host galaxy and its nearest neighbor, respectively, where the projected virial radius is 
defined as the projected radius which encloses the mean mass density as high as $200$ times the critical density of the 
universe. 

For an isolated host galaxy with absolute $r$-band magnitude of $M_{r}$, the candidates of its satellites satisfy the following 
two conditions:  First, the absolute $r$-band magnitude is larger than $M_{r}+1$. Second, the difference in the radial velocity 
between a satellite candidate and its host, $\Delta v_{s}$, is smaller than the escape velocity of a satellite, $\Delta v_{s,c}$. 
Taking into account the dependence of the escape velocity of a satellite on the mass of its host galaxy and assuming that 
the constant mass-to-light ratio of an elliptical host galaxy is twice that of an spiral galaxy, \citet{HP10} derived an equation of 
$\Delta v_{s,c}\equiv \Delta v_{c}\ 10^{-2(M_{r}+21)/15}$ where the amplitude $\Delta v_{c}$ has the value of 
 $900$km/s and $600$km/s for the elliptical and the spiral galaxies, respectively. These values of $\Delta v_{c}$ were 
 empirically determined from the work of \citet{PC09} who studied the frequency distribution of the differences in the radial 
 velocities between the SDSS galaxy pairs. In the works of \citet{PC09} and \citet{HP10}, the elliptical and spiral galaxies were 
 classified according to the morphology segregator that had been developed by \citet{PC05} based on the gradients of 
 the galaxy colors.

The true satellites of an isolated host galaxy are identified among the candidates as the ones whose separation distances 
from the isolated host, $r_{s}$, in the projected plane of sky are less than some threshold, $r_{sc}$, given 
as  \citep{HP10} 
%%%%%%%%%%%%%%%%%%%%%%%%%%%%%%%%%%%%%%%%%%%%%%%%%%%%%%%%%%
\begin{equation}
\label{eqn:satbound}
r_{sc} = {\rm min}\left\{\frac{r_{n}R_{\rm v}}{R_{\rm v}+R_{\rm vn}},\  r_{n}-R_{\rm vn}\, ,\ 3\,h^{-1}{\rm Mpc}\right\}\, .
\end{equation}
%%%%%%%%%%%%%%%%%%%%%%%%%%%%%%%%%%%%%%%%%%%%%%%%%%%%%%%%%%
For the detailed description of how to identify the isolated galactic systems, the readers are referred to 
\citet{PC09} and \citet{HP10}. 

\begin{figure}[tb]
\includegraphics[scale=0.8]{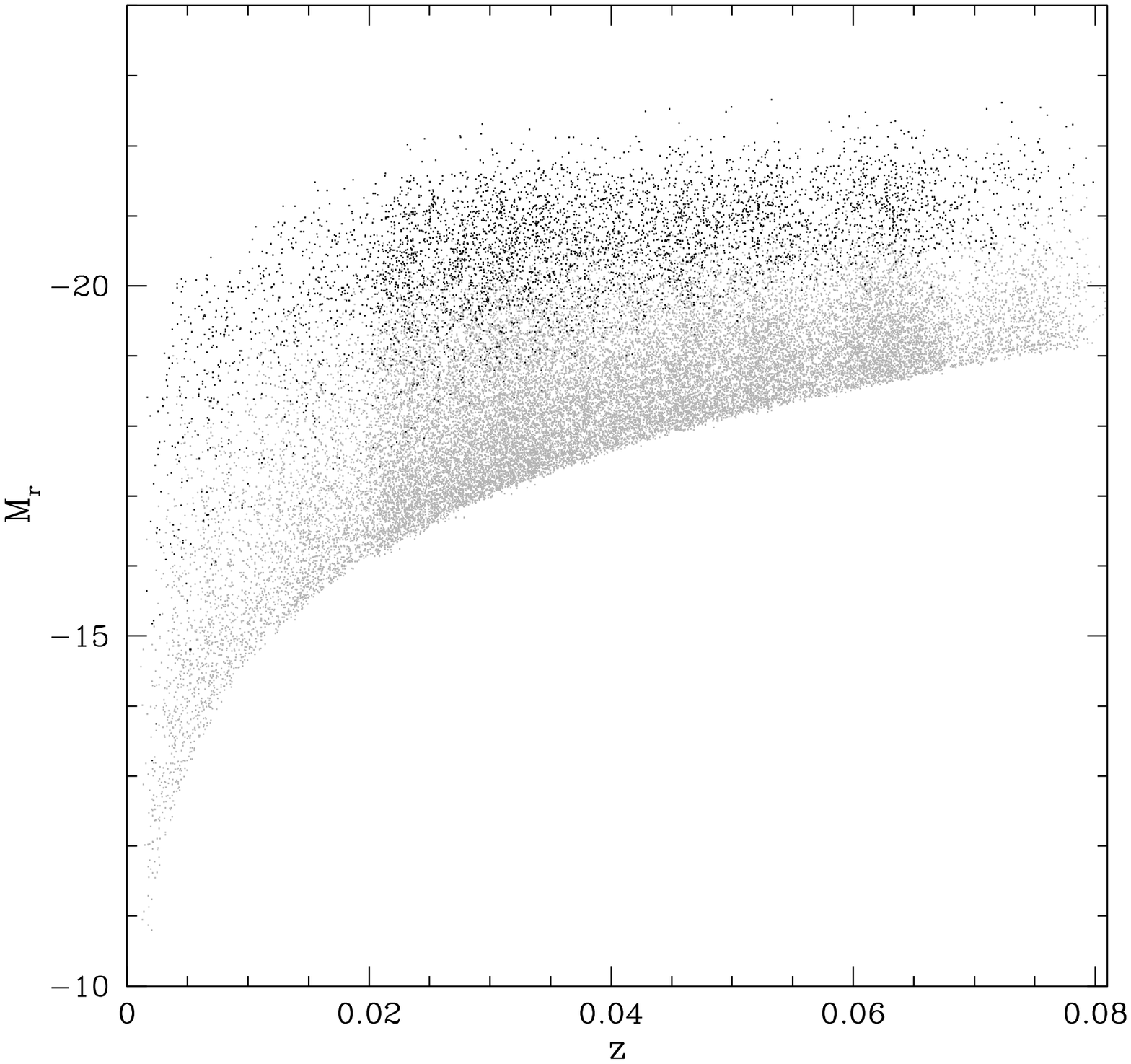}
\caption{Absolute $r$-band magnitudes of the isolated host galaxies (black dots) and their satellites (grey dots) 
from the SDSS DR7 as a function of the redshift. For visibility, only those isolated galactic systems with five or more 
satellites are shown. }
\label{fig:mag}
\end{figure}

\begin{figure}
\includegraphics[scale=0.8]{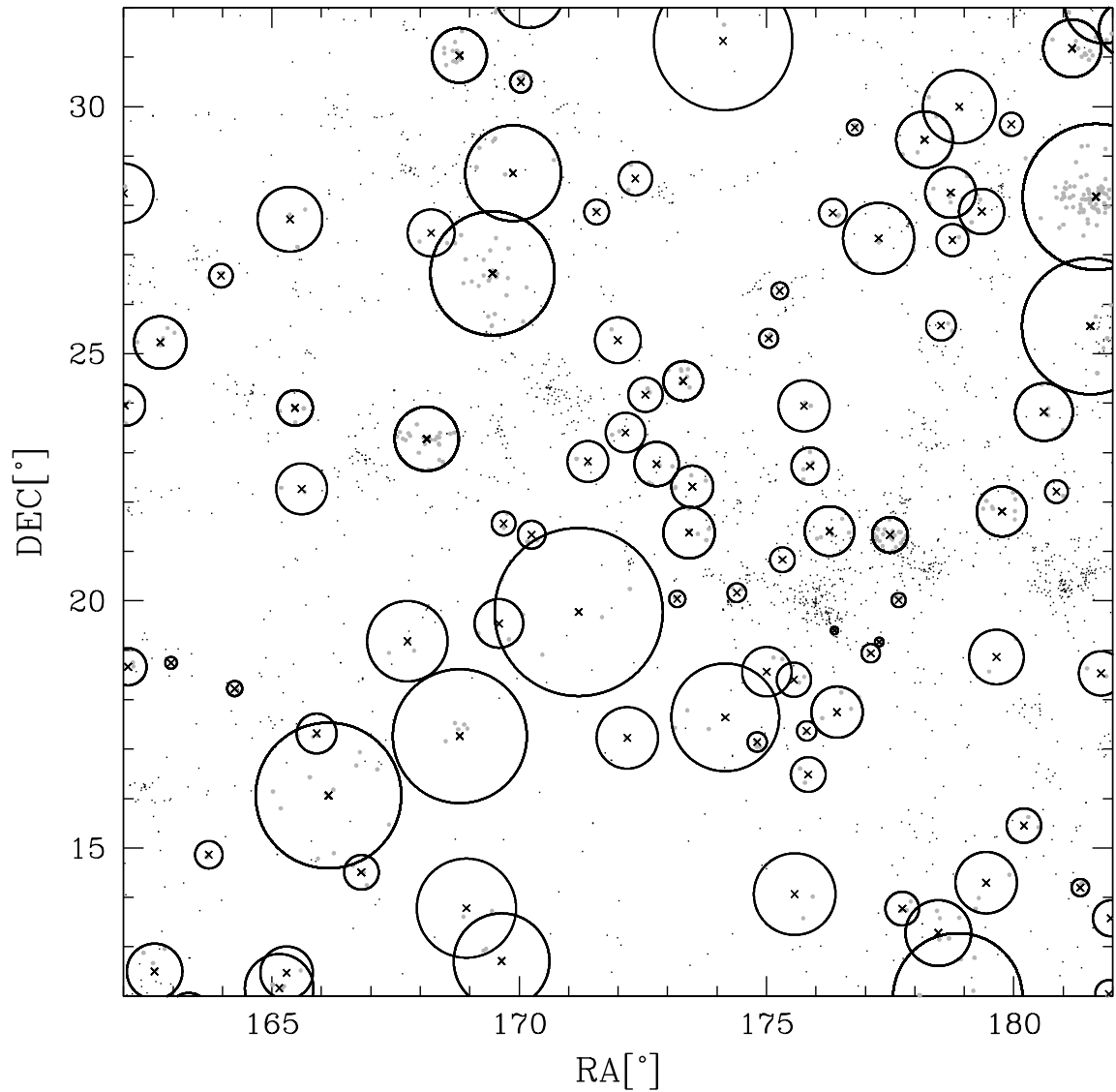}
\caption{Illustration of how to identify the satellites (grey dots)  around the isolated host galaxies (cross) in the 
plane of sky spanned by the right ascension (RA) and declination (DEC) in unit of degree at redshifts of $0.02\le z\le 0.03$. 
The radii of the  circles around the isolated host galaxy represent the maximum separation distances to their satellites. 
The apparent overlapping of some circles with the neighbour circles around the host galaxies are due to the projection effect.}
\label{fig:circle}
\end{figure}

A total of $10646$ SDSS galaxies are selected as the isolated hosts having one or more satellites. 
The ranges of the absolute $r$-band magnitudes, the projected virial radius and the satellite number ($N_{s}$) of the 
selected hosts are found to be $-22.66\le M_{r}\le -12.85$, $0.027\le R_{\rm v}\le 0.687$ and $1\le N_{s}\le 18$, 
respectively. Figure \ref{fig:mag} plots the absolute $r$-band magnitudes of the isolated host galaxies (black dots) and their 
satellites (grey dots) from the selected sample of the SDSS DR7 as a function of redshift. 
Figure \ref{fig:circle} illustrates the configuration of the satellites (grey dots) identified around the isolated host galaxies 
(black cross), whose declination (DEC), right ascension (RA) and redshifts are in the ranges of 
$12^{\rm o}\le$DEC$\le 32^{\rm o}$, $162^{\rm o}\le$RA$\le 182^{\rm o}$, and $0.02\le z\le 0.03$, respectively.  
The radius of the circle around each isolated host galaxy corresponds to the threshold separation 
distance, $r_{sc}$, in Equation (\ref{eqn:satbound}). The apparent overlapping among the neighbouring circles around 
the host galaxies are merely due to the projection effect. For the visibility, only those galactic systems having five or 
more satellites are shown in Figures \ref{fig:mag}-\ref{fig:circle}. 

To determine the principal axes of the local velocity shears at the positions of the selected host galaxies, we utilize the density 
and velocity fields reconstructed by \citet{wang-etal12} from the galaxy group catalog based on 
the SDSS DR7 \citep{blanton-etal05,sdssdr7}. \citet{wang-etal12} estimated the masses of the galaxy groups according 
to the prescriptions of \citet{yang-etal07} which is based on the ranking of the stellar masses of the galaxies
and a varying mass-to-light ratio model. For a detailed description of the halo-mass estimator, see \citet{yang-etal07}. 
Selecting only those galaxy groups with $M\ge 10^{12}\,h^{-1}M_{\odot}$ located in the "survey volume" which spans the 
redshift range of $0.01\le z\le 0.12$, and taking into account the redshift distortion effect on the positions of the selected 
galaxy groups, \citet{wang-etal12} reconstructed the density field $\delta({\bf x})$ with the help of the halo tracing algorithm.  
The reconstructed density field $\delta({\bf x})$ was defined on the $494\times 892\times 499$ cubic cells comprising the 
survey volume, where each cell has a linear size of $l_{c}\approx 0.71\,h^{-1}$Mpc and its position ${\bf x}$ is measured in 
the equatorial coordinate systems

Transforming the density field $\delta({\bf x})$ into the Fourier space and assuming that the peculiar velocity field 
${\bf v}({\bf x})$ is curl free and thus related to the density field as ${\bf v} \propto \nabla\nabla^{-2}\delta$, \citet{wang-etal12} 
determined the Fourier amplitudes of the filtered peculiar velocity field $\tilde{\bf v}({\bf k})$ from those of the density field 
$\tilde{\delta}({\bf k})$ as $\tilde{\bf v}({\bf k}) \propto ({\bf k}/k^{2})\,\tilde{\delta}\,({\bf k})W(kR_{f})$ where ${\bf k}$ is the 
Fourier-space wave vector with magnitude of $k\equiv \vert{\bf k}\vert$ and $W(kR_{f})$ is the Fourier-space Gaussian window 
function with a filtering radius of $R_{f}$. The peculiar velocity field smoothed on the scale of $R_{f}$ was reconstructed via the inverse 
Fourier transformation of $\tilde{\bf v}$. 

As stated in \cite{wang-etal12}, the accuracy of the reconstructed density and peculiar velocity fields depends on the filtering 
scale.  On small scales $R_{f}< 2l_{c}$, the reconstructed fields are expected to be inaccurate due to the limited spatial  
resolution. On large scales, the reconstruction would also fail to yield a good approximation due to the 
presence of the boundary effect. 
 \citet{wang-etal12} claimed that when $R_{f}\approx 2\,h^{-1}$Mpc,  the reconstruction worked well in the 
low-redshift inner cells (close to the center) that occupy the $66\%$ of the survey volume. The portion of the survey volume 
where the density field is reconstructed with high accuracy would diminish as the filtering scale $R_{f}$ increases. 
Since we will consider larger scales of $R_{f}> 2\,h^{-1}$Mpc to explore the shear-satellite alignments in section 
\ref{sec:analysis}, we have to restrict our analysis to the innermost regions of the survey volume to 
secure the reliability of the reconstructed velocity shear fields. 

We find that the isolated galactic systems in our sample can be covered by the innermost subvolume of $180\,h^{-3}$Mpc$^{3}$ 
consisting of $256^{3}$ cells whose equatorial positions, $(x,\ y,\ z)$, range as 
$-182.29\le x/(h^{-1}{\rm Mpc}) \le -1.49,\ -90.05\le y/(h^{-1}{\rm Mpc})\le 90.74,\ -22.73\le z/(h^{-1}{\rm Mpc})\le 188.07$.  
Here, the innermost subvolume represents the regions close to the center of the survey volume corresponds to the redshift 
range of $0 \le z \le 0.085$. The velocity shear field defined $\Sigma_{ij}({\bf x})\propto \partial_{i}v_{j}+\partial_{j}v_{i}$ 
\citep{hoffman-etal12} has been linearly reconstructed on this subvolume \citep{lee-etal14}.  Here we calculate the Fourier 
amplitudes of the velocity shear field, $\tilde{\Sigma}_{ij}({\bf k})$, directly from the Fourier amplitudes of the density field 
$\tilde{\delta}({\bf k})$ as $\tilde{\Sigma}_{ij}\propto ({k}_{i}k_{j}/k^{2})\,\tilde{\delta}({\bf k})\,W(kR_{f})$ and finally 
obtain the velocity shear field, $\Sigma_{ij}({\bf x})$, smoothed on the scale of $R_{f}$ via the inverse Fourier transformation. 

Note that the {\it linearly reconstructed} velocity shear field is essentially the same (apart from the overall amplitude) as 
the tidal shear field $T_{ij}$ defined as the second derivative of the gravitational potential as 
$T_{ij}\propto \partial_{i}\partial_{j}\phi$ since the gravitational potential $\phi$ is related to the density field as  
$\phi\propto \nabla^{-2}\delta$. In other words, the directions of the principal axes of $\Sigma_{ij}$ are the 
same as those of $T_{ij}$.  On small scales where the real velocity field should no longer be assumed to be curl free, 
the linearly reconstructed velocity field would fail to approximate well the real velocity field. However, 
\citet{libeskind-etal14a} showed by $N$-body simulations that the vorticity quickly vanishes as 
the filtering scale increases and that the directions of the principal axes of the real velocity shear field do not change 
sensitively with the smoothing scales over a wide range, which implied that the linearly reconstructed 
velocity shear field should be a good approximation to the real velocity field on large scales. 

\section{ALIGNMENTS BETWEEN THE GALACTIC SATELLITES AND THE LARGE-SCALE VELOCITY SHEARS}
\label{sec:analysis}

Using information on the redshifts and equatorial coordinates of the host and its satellites in each of the isolated galactic 
systems, we determine the three dimensional (comoving) positions, ${\bf r}_{3d,s}$, of the satellites relative to the host. At the 
cell to which the position of each isolated host belongs, we diagonalize the velocity shear tensor $(\Sigma_{ij})$ through the 
similarity transformation to evaluate its three eigenvalues (say, $\lambda_{1},\ \lambda_{2},\ \lambda_{3}$ in a decreasing order) 
and determine the corresponding three eigenvectors (say, ${\bf e}_{3d,1},\ {\bf e}_{3d,2},\ {\bf e}_{3d,3}$) as the major, 
intermediate and minor principal axes of the velocity shear tensor. 

\begin{figure}
\includegraphics[scale=1.0]{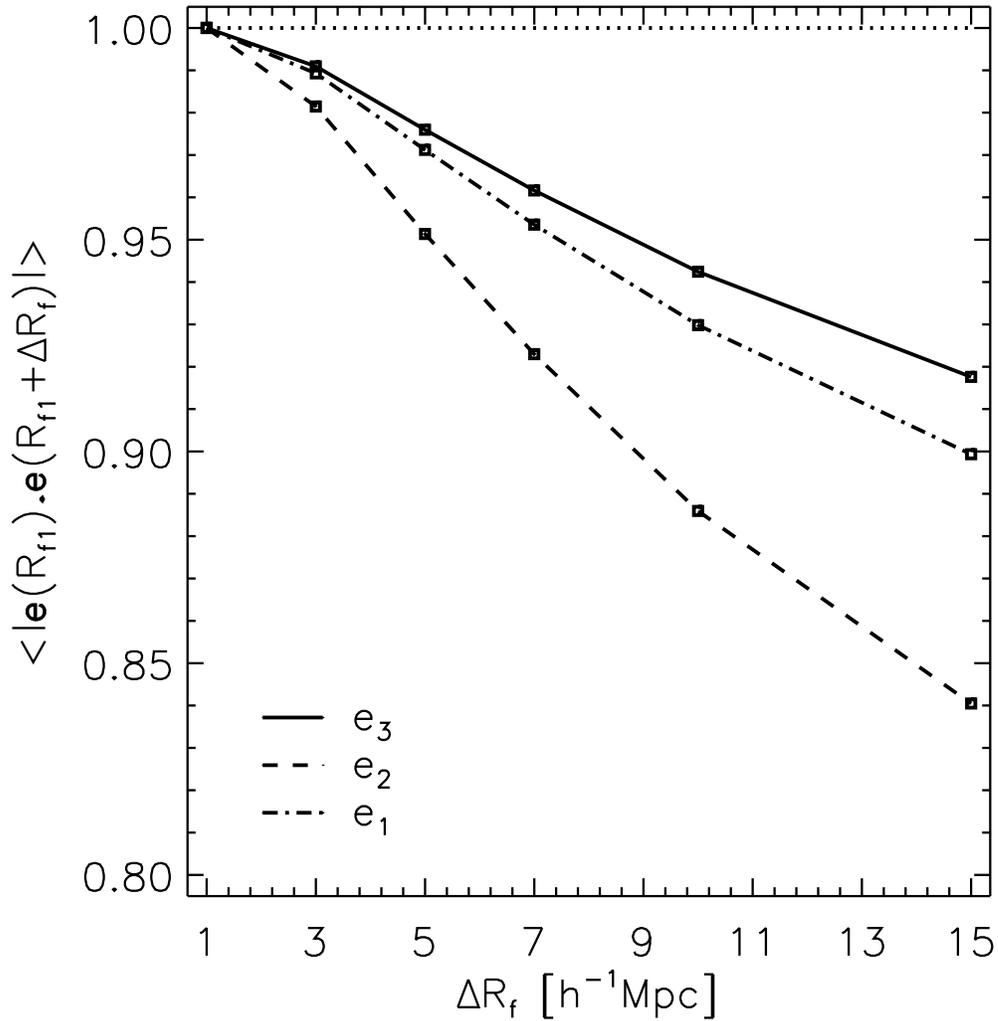}
\caption{3D cross-correlations of the principal axes of the velocity shear fields between two scales $R_{f1}$ and $R_{f2}$  
as a function of the scale difference $\Delta R_{f}=R_{f2}-R_{f1}$ with $R_{f1}$ set at $1 h^{-1}$Mpc. The solid, dashed and 
dot-dashed line corresponds to the cross-correlations of the minor, intermediate and major principal axes of the velocity shear 
fields.}
\label{fig:cross}
\end{figure}

To see how the directions of the principal axes of the local velocity shears estimated at the positions of the host galaxies change 
with the filtering scale, we evaluate the average of the cross-correlations between the principal axes of the local velocity shears 
smoothed on two different filtering scales,  $R_{f1}$ and $R_{f2}$, as 
$\langle\vert{\bf e}_{3d,i}(R_{f1})\cdot{\bf e}_{3d,i}(R_{f2})\vert\rangle$ for $i=1,\ 2,\ 3$. 
Figure \ref{fig:cross} plots the cross-correlations as a function of the difference, $\Delta R_{f}\equiv R_{f2}-R_{f1}$, with $R_{f1}$ 
set at $1\,h^{-1}$Mpc. 
As can be seen, the cross-correlations decrease quite slowly as $\Delta R_{f}$ increases.  Especially the minor principal 
axes exhibit the slowest decrement among the three axes, showing less than $5\%$ change as $\Delta R_{f}$ 
varies over a wide range from $1\,h^{-1}$Mpc to $10\,h^{-1}$Mpc.  As mentioned in section \ref{sec:data}, this result 
agrees well with the numerical finding of \citet{libeskind-etal14a} who demonstrated by $N$-body simulations 
that the orientations of the principal axes of the velocity shear field are quite robust against the changes of the filtering scale.  
It is, however, worth recalling that \citet{libeskind-etal14a,libeskind-etal14b} utilized the real velocity shear field while here we 
deals with the linearly reconstructed one. 

Given that the positions of the galaxies as well as the directions of the principal axes of the velocity shear tensors determined in 
the redshift space are likely to suffer from systematic errors, we examine whether or not the distributions of ${\bf r}_{3d,s}$ 
and $\{{\bf e}_{3d,i}\}_{i=1}^{3}$ in each galactic system are isotropic relative to the line-of-sight direction to its host galaxy, say, 
$\hat{\bf s}$. First, we measure the alignment between the relative positions of the satellites ${\bf r}_{3d,s}$ and the unit vector in the line of sight direction as $\mu_{los}\equiv \vert {\bf r}_{3d,s}\cdot\hat{\bf s}\vert/\vert{\bf r}_{3d,s}\vert$ and take the average 
over all of the selected galactic systems. If there were no systematic errors in the measurement of the positions of the satellites 
in the line-of-sight directions,  the average value of $\mu_{los}$ would be close to $0.5$. 
However, the actual average value of $\mu_{los}$ has turned out to be $0.772\pm 0.003$ which significantly exceeds the value 
of $0.5$. 

Similarly, we compute the average of the alignments between $\{{\bf e}_{3d,i}\}_{i=1}^{3}$ and $\hat{\bf s}$ as 
$\mu_{los,e_{i}}\equiv \langle\vert ({\bf e}_{3d,i}\cdot\hat{\bf s})/\vert{\bf e}_{3d,i}\vert\rangle$ for $i=1,\ 2,\ 3$. 
The average values are found to be 
$\mu_{los,e_3}=0.558\pm 0.003,\ \mu_{los,e_2}=0.494\pm 0.003,\ \mu_{los,e_1}=0.441\pm 0.003$, when the filtering scale 
for the velocity shear field is set at $R_{f}=5\,h^{-1}$Mpc.
These results reveal that the principal axes of the local velocity shears measured at the positions 
of the host galaxies are not isotropically oriented relative to the line-of-sight directions but the minor ones appear to be 
preferentially aligned with $\hat{\bf s}$.  Therefore, if the alignments between ${\bf r}_{3d,s}$ and ${\bf e}_{3d,i}$ are measured 
in the three dimensional redshift space,  the result would be contaminated by the presence of a spurious signal induced by the 
{\it extrinsic} alignments of both of ${\bf r}_{3d,s}$ and ${\bf e}_{3d,i}$ with the line of sight directions.  

\begin{figure}[tb]
\includegraphics[scale=0.7]{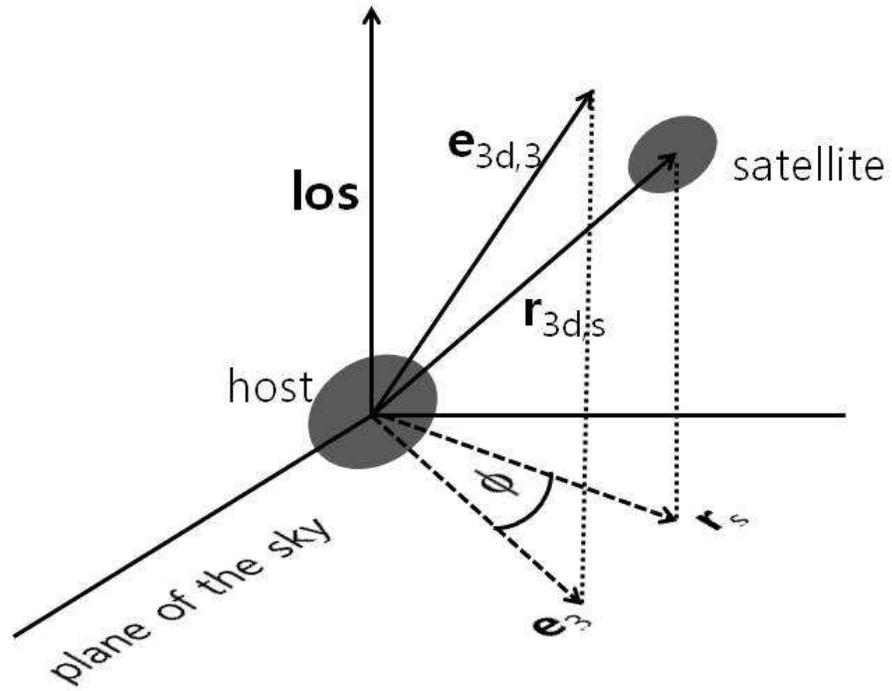}
\caption{Illustration of the alignment angle, $\phi$, between the projected two dimensional position of a satellite around its 
isolated host and the projected direction of the minor principal axis of a local velocity shear tensor in the plane of the sky 
perpendicular to the line-of-sight (los) direction to the host.}
\label{fig:conf}
\end{figure}

To eliminate a spurious signal and detect a true signal of the {\it intrinsic} alignments between the principal axes of the velocity 
shear field and the positions of the satellites, we project them onto the plane of sky, as done in \citet{lee-etal14}. 
From here on, ${\bf r}_{s}$ and $\{{\bf e}_{1},\ {\bf e}_{2},\ {\bf e}_{3}\}$ represent the two dimensional satellite positions and 
the principal axes of the local velocity shear tensors projected onto the plane of sky perpendicular to $\hat{\bf s}$, respectively. 
We define the {\it alignment angle}, $\phi_{i}$, of each selected system as the mean value of the angles between the projected 
positions of the satellites relative to their host and the projected $i$-th principal axes of the local velocity shear tensor: 
%%%%%%%%%%%%%%%%%%%%%%%%%%%%%%%%%%%%%%%%%%%%%%%%%%%%%%%%%%%%%%%
\begin{equation}
\label{eqn:mu}
\phi_{i} \equiv \frac{1}{N_{s}}\sum_{\alpha}\cos^{-1}
\left(\frac{\vert{\bf r}_{s}^{\alpha}\cdot{\bf e}_{i}\vert}
{\vert{\bf r}_{s}^{\alpha}\vert\vert{\bf e}_{i}\vert}\right) \, , \qquad {\rm for}\quad i=1,\ 2,\ 3\, ,
\end{equation}
%%%%%%%%%%%%%%%%%%%%%%%%%%%%%%%%%%%%%%%%%%%%%%%%%%%%%%%%%%%%%%
where $N_{s}$ is the number of the satellites belonging to a given system, ${\bf r}_{s}^{\alpha}$ is the two dimensional 
position of the $\alpha$-th satellite.  If the satellites were isotropically distributed with respect to the $i$-th principal axes 
of the velocity shear field, then $\phi_{i}$ would be close to $45^{\rm o}$ for $i=1,\ 2,\ 3$. If $\phi$ is lower (higher) 
than $45^{\rm o}$, then it should imply a signal of alignment (anti-alignment).   
Figure \ref{fig:conf} illustrates how the alignment angle is measured between the projected position of a satellite 
and the projected minor principal axis of a local velocity shear tensor at the location of a host galaxy  
in the plane of the sky.  

We determine the alignment angle $\phi_{i}$ for each galactic system and take the ensemble average 
of $\phi_{i}$ over all galactic systems for four different cases of the filtering scale: $R_{f}=1,\ 5,\ 10,\ 15\, h^{-1}$Mpc, 
the results of which are shown Table \ref{tab:rf}. The error $\sigma_{\phi}$ represents the standard deviation in the 
measurement of $\langle\phi_{i}\rangle$ calculated as $\sigma_{\phi}\equiv \sqrt{\langle\Delta^{2}\phi_{i}\rangle/(N_{sys}-1)}$ 
where $N_{sys}$ denotes the number of the selected galactic systems. 
As can be seen, the projected positions of the satellites of the isolated galactic systems are aligned with the projected minor 
principal axes and anti-aligned with the projected major principal axes of the local velocity shears. 
Although the strength of the alignment signal seems to decrease as $R_{f}$ increases, the alignment signal does 
not disappear even when the filtering scales reaches up to $15\,h^{-1}$Mpc.
At this stage, an intriguing question that arises is at what filtering scale the alignment signal would disappear. The current 
analysis, however, cannot provide an answer to this question since we cannot consider an arbitrarily large filtering scale 
because of the fact that the reconstructed velocity shear field becomes more and more inaccurate as $R_{f}$ increases  
\citep{wang-etal12}.  

\section{DEPENDENCE OF THE ALIGNMENT SIGNAL ON THE GALAXY PROPERTIES}
\label{sec:depend}

The results presented in section \ref{sec:analysis} provides a clear  observational evidence that the spatial locations of the 
galactic satellites around the isolated hosts are preferentially aligned with the minor principal axes of the local velocity 
shear field.  Given that the projection procedure must have played a role in weakening the alignment signal, it is expected that 
the true alignment tendency must be stronger than detected in the current analysis.  Now, we would like to investigate if and 
how the alignment strength depends on the properties of the galactic systems. 

Let us first see how the alignment angles depend on the distances between the satellites and the hosts. 
For each selected galactic system,  we divide the satellites into two subsamples according to their projected distances $r_{s}$ 
from their hosts measured in the plane of the sky. One subsample consists of those satellites with $r_{s} \le R_{v}$ while the 
other has the satellites with $r_{s} > R_{\rm v}$, where $R_{\rm v}$ denotes the projected virial radius of their host. 
The alignment angles, $\{\phi_{i}\}_{i=1}^{3}$, of each system are measured twice by using the two subsamples separately. 
The averages of them over all of the selected galactic systems are also separately taken. 

The first and second rows of Table \ref{tab:satellite} show the averages of the alignment angles, $\{\langle\phi_{i}\rangle\}_{i=1}^{3}$, 
for the two cases that the values of $\{\phi_{i}\}_{i=1}^{3}$ of each galactic system are measured by using only those satellites whose 
separation distances $r_{s}$ from their host are smaller and larger than the projected virial radius, $R_{\rm v}$, respectively. 
The filtering scale of the velocity shear field is set at $R_{f}=5\,h^{-1}$Mpc from here on. 
Although separating the satellites into the two subsamples resulted in larger errors, one can still see a clear alignment 
tendency for the case of the satellites with $r_{s}>R_{\rm v}$. Whereas, the satellites with $r_{s}\le R_{\rm v}$ appear to be almost 
isotropically distributed with respect to the principal axes of the local velocity field. 
This result implies that the alignment tendency between the spatial locations of the satellites and the minor principal axes of 
the local velocity shear field diminishes as the satellites enter the virial radii of their hosts where the non-linear effect is 
expected to be dominant.   

Next, we investigate how the alignment signal depends on the absolute $r$-band magnitudes of the satellites. 
Separating the brighter satellites of each galactic system with $M_{r}\le M_{r0}$ from the fainter ones with $M_{r}>M_{r0}$ 
where $M_{r,0}=-18.2$ is the median value of the absolute $r$-band magnitudes of the satellites, the values of $\{\phi_{i}\}$ 
of each galactic system are measured twice by using the two subsamples separately. 
The third and fourth rows of Table \ref{tab:satellite} shows $\{\langle\phi_{i}\rangle\}_{i=1}^{3}$ for the two cases that the values 
of $\{\phi_{i}\}_{i=1}^{3}$ of each galactic system are measured by using the brighter and fainter satellites, respectively. 
Although the relatively large errors prevent us from finding a statistically significant signal, the result hints at how the 
alignment strength depends on $m_{r}$: the brighter galaxies are more strongly aligned with the minor principal axes of the 
local velocity shear tensors than the fainter counterparts,  which is in fact consistent with the previous numerical results 
\citep[][and references therein]{libeskind-etal14b} that the more massive satellites tend to be more preferentially aligned with the 
minor principal axes of the local velocity shear field. 

We also investigate whether or not the alignment strength depends on the morphology of the satellites. Using 
the morphology segregator algorithm suggested by \citet{PC05}, we classify the satellites of each galactic system 
into the ellipticals and the spirals.  Then, the alignment angles of each galactic system are measured twice by 
using separately the elliptical and the spiral satellites, the result of which is displayed in the fifth and sixth rows of 
Table \ref{tab:satellite}. 
As can be seen, although the errors are large, the alignment tendency can be found for the case of the 
spiral satellites  but no alignment signal for the case of the elliptical satellites.  

Our interpretation of the results shown in Table \ref{tab:satellite} is that the strength of the alignment between the 
relative positions of the galactic satellites and the minor principal axes of the velocity shear field depends on the formation and 
accretion epochs of the galactic satellites. The satellites have accreted onto the hosts in the directions aligned with the minor 
principal axes of the linear velocity shear field. After they become bound to the host system, the nonlinear modifications 
gradually weeken the tendency of the alignments between their positions and the minor principal axes of the velocity 
shears.  The earlier the satellites have formed and accreted onto their hosts, the less strong the alignments of their locations 
with the minor principal axes of the velocity shear fields are. Comparing this result of our analysis with those 
from the previous works \citep{bailin-etal08,AB10,dong-etal14} which found that the elliptical galaxies are more strongly aligned 
with the elongated axes of the host galaxies,  we suspect that the alignments between the locations of the satellites and the 
shapes of their hosts should be modulated by the nonlinear interaction between the satellites and the hosts after the accretion 
of the satellites. 
 
Now, we would like to see if and how the alignment strength depends on the properties of the host galaxies as well as 
the surrounding environment.  We first investigate dependence of the alignment tendency on the host size. The median 
value of the projected virial radii of the selected isolated hosts is found to be $0.285\,h^{-1}$Mpc. Determining the 
alignment angle $\{\phi_{i}\}_{i=1}^{3}$ of each galactic system, we calculate the ensemble average, 
$\{\langle\phi_{i}\rangle\}_{i=1}^{3}$ over those systems whose hosts have $R_{\rm v}$ larger than the median value. 
We repeat the same but only over those with $R_{\rm v}$ equal to or less than the median value. The first and second rows 
of Table \ref{tab:host} show the result, indicating that the alignment strength is insensitive to the sizes of the host galaxies. 

We also examine the dependence of the shear-satellite alignments on the absolute $r$-band magnitudes of the hosts. The median 
value of the absolute $r$-band magnitudes of the selected isolated hosts is found to be $-20.3$. We determine 
$\{\langle\phi_{i}\rangle\}_{i=1}^{3}$ separately over those systems whose hosts have $M_{r}\le -20.3$ and over $M_{r}>-20.3$. 
The results shown in the third and fourth rows of Table \ref{tab:host} indicate a trend of almost independence of  the 
shear-satellite alignment strength on the absolute $r$-band magnitudes of the host galaxies. 
 
To see if the strength of the shear-satellite alignment depends on the morphology of the hosts, we calculate  
$\{\langle\phi_{i}\rangle\}_{i=1}^{3}$ separately over those systems with spiral hosts and repeat the same calculation 
but over those whose hosts are ellpiticals, the results of which are shown in the fifth and sixth rows of Table \ref{tab:host}. 
As can be seen, there is a tendency that the galactic satellites belonging to the spiral hosts show higher strength of the 
alignments with the minor principal axes of the velocity shear tensors. This tendency is consistent with the picture  
that the non-linear effect breaks the shear-satellite alignments after the formation and accretion of the satellites 
since the elliptical hosts are believed to have formed earlier and thus their satellites were more severely affected by the 
nonlinear effect.

Using the web-classification scheme suggested by \citet{hahn-etal07}, we determine in which web environment 
each galactic system reside and find that most of the systems are located either in the filament 
where $\lambda_{1}\ge\lambda_{2}\ge 0$ and $\lambda_{3}<0$ 
or in the sheet where $\lambda_{1}\ge 0$ and $\lambda_{2}<0$.  It is worth mentioning here why we used the web-classification 
scheme proposed by \citet{hahn-etal07} rather than that of \citet{hoffman-etal12}. The difference between the two schemes 
is that the web classification of \citet{hoffman-etal12} adopted the positive threshold of $\lambda_{th}>0$ while in the analysis 
of \citet{hahn-etal07} the threshold value was set at zero $\lambda_{th}=0$. In the currently analysis, we have used the {\it linearly 
reconstructed} velocity shear field which is same as the tidal shear field and thus adopt the zero threshold value of 
$\lambda_{th}=0$. In fact, even in \citet{hoffman-etal12} this zero threshold value was used to classify the web environment 
for the investigation of the universality of the velocity shear effect. 

Among the $10646$ selected galactic systems, a total of $4975$ ($4422$) systems are found to be in the filament (sheet). 
The rest of the selected systems are found to reside either in the knot or in the void.
After calculating $\{\phi_{i}\}_{i=1}^{3}$ of each galactic system, we take its ensemble average over 
only those systems located in the filament. We repeat the same but over those systems located in the sheet. 
We do not consider the galactic systems residing in the void or in the knot since the evaluation of 
$\{\langle\phi_{i}\rangle\}_{i=1}^{3}$ would suffer from small number statistics for those cases.
The seventh and eighth rows of Table \ref{tab:host} show how the ensemble averages of the alignment angles are 
different between the two cases that the galactic systems are embedded in the sheet and in the filament environments. 
As can be seen, there is almost no difference between the two cases, which is consistent 
with the numerical prediction of \citet{libeskind-etal14b} based on a high-resolution $N$-body simulation.

\section{SUMMARY AND DISCUSSION}\label{sec:con}
 
Our work has been done in the light of two recent literatures. The first one is \citet{libeskind-etal14b} which claimed with the help 
of a $N$-body experiment that the local velocity shear field dictates the anisotropic infalls of the satellites onto their hosts. The 
second one is \citet{lee-etal14} which found an observational evidence for the anti-alignments of the locations of the cluster 
galaxies  with the major principal axes of the local velocity shears by analyzing the extended Virgo cluster catalog 
\citep{kim-etal14}.  

Using a sample of the isolated galactic systems in the redshift range of $0< z < 0.08$ from the SDSS DR7 
and the velocity shear field linearly reconstructed in the local volume where the isolated galactic systems are located, we have 
analysed the spatial distributions of the galactic satellites in the principal frame of the local velocity shear tensor and 
found a clear signal of the alignment between the locations of the galactic satellites and the minor principal axes of the 
velocity shear field projected onto the plane of sky. The projection procedure has been conducted to avoid any false alignment 
signal produced by systematic errors in the measurements of the positions of the galactic satellites as well as the principal axes 
of the velocity shear field in the redshift space. The satellite-shear alignment tendency has been shown to be stable against the 
changes of the filtering scale. 

Investigating the dependence of the alignment tendency on the properties of the satellites, we have found stronger 
alignment signals for the cases that the galactic satellites are spirals, more luminous, located at distances larger than the 
projected virial radii of their hosts and belonging to the spiral hosts.  
Given these results, we have suggested that the strength of the satellite-shear alignment should depend most 
sensitively on the formation and accretion epochs of the galactic satellites. The earlier the satellites formed and accreted onto 
their host galaxies, the less strong the satellite-shear alignments are since the nonlinear modification after the formation 
and accretion of the satellites would play a role in diminishing the satellite-shear alignment tendency.  
Given the results of the previous works \citep[e.g,][]{SL04,bailin-etal08,AB10,dong-etal14} which found that 
the preferential alignments with the elongated axes of the host galaxies were exhibited only by the elliptical satellites 
but not by the spiral counterparts,  we speculate that the alignments of the locations of the satellites with the elongated axes 
of the hosts should be modulated dominantly by the non-linear effect.

We have also noted a trend that the strength of the satellite-shear alignment does not vary with the sizes and 
luminosities of the host galaxies and the web environment, which is in line with the numerical prediction of \citet{libeskind-etal14b} 
based on a $N$-body simulation. However, due to the very low statistical significance, the observed signals from our analysis 
cannot not provide a convincing direct evidence for the numerical claim of \citet{libeskind-etal14b} and thus it is still inconclusive 
whether or not the velocity shear has a truly universal effect on the satellite infall motions. It would require a larger dataset to 
detect a signal with high statistical significance.

In addition, to confirm the universality of the velocity shear effect on the satellite infall motions, the improvement of the analysis should 
be made in the following two directions.
First, it will be essential to develop an algorithm for the reconstruction of the nonlinear velocity shear field. Although the linearly 
reconstructed velocity shear field is believed to be a good approximation to the real one on large scales  
\citep{lee-etal09,libeskind-etal14a}, this linear approximation would not be valid on small scales where the 
nonlinear effect becomes significant and the velocity field develops vorticity. 
Second, as in the numerical work of \citet{libeskind-etal14b}, it will be necessary to determine the true infall directions of the satellites 
from observations. Although the direction of the spatial position of a satellite and that of its infall motion are expected to be strongly 
aligned with each other, they should not be equal. 

\acknowledgments

Funding for the SDSS and SDSS-II has been provided by the Alfred P. Sloan Foundation, 
the Participating Institutions, the National Science Foundation, the U.S. Department of 
Energy, the National Aeronautics and Space Administration, the Japanese Monbukagakusho, 
the Max Planck Society, and the Higher Education Funding Council for England. The 
SDSS Web Site is http://www.sdss.org/. 

The SDSS is managed by the Astrophysical Research Consortium for the Participating 
Institutions. The Participating Institutions are the American Museum of Natural History, 
Astrophysical Institute Potsdam, University of Basel, University of Cambridge, Case 
Western Reserve University, University of Chicago, Drexel University, Fermi lab, the 
Institute for Advanced Study, the Japan Participation Group, Johns Hopkins University, 
the Joint Institute for Nuclear Astrophysics, the Kavli Institute for Particle 
Astrophysics and Cosmology, the Korean Scientist Group, the Chinese Academy of Sciences 
(LAMOST), Los Alamos National Laboratory, the Max-Planck-Institute for Astronomy (MPIA), 
the Max-Planck-Institute for Astrophysics (MPA), New Mexico State University, Ohio State 
University, University of Pittsburgh, University of Portsmouth, Princeton University, 
the United States Naval Observatory, and the University of Washington. 

We thank an anonymous referee for valuable suggestions which helped us significantly improve the original manuscript.
J.L. thanks H. Wang for providing the dataset of the density and peculiar velocity field. This work was supported by the 
research grant from the National Research Foundation of Korea to the Center for Galaxy Evolution Research  
(NO. 2010-0027910). 
J.L. also acknowledges the financial support by the Basic Science Research Program through the National Research 
Foundation of Korea (NRF) funded by the Ministry of Education (NO. 2013004372).

%%%%%%%%%%%%%%%%%%%%%%%%%%%%%%%%%%%%%%%%%%%%%%%%%%%%%%%%%%%
\clearpage
%%%%%%%%%%%%%%%%%%%%%%%%%%%%%%%%%%%%%%%%%%%%%%%%%%%%%%%%%%%
%%%%%%%%%%%%%%%%%%%%%%%%%%%%%%%%%%%%%%%%%%%%%%%%%%%%%%%%%%%
\clearpage
%%%%%%%%%%%%%%%%%%%%%%%%%%%%%%%%%%%%%%%%%%%%%%%%%%%%%%%%%%%
%%%%%%%%%%%%%%%%%%%%%%%%%%%%%%%%%%%%%%%%%%%%%%%%%%%%%%%%%%%
\clearpage
%%%%%%%%%%%%%%%%%%%%%%%%%%%%%%%%%%%%%%%%%%%%%%%%%%%%%%%%%%%
%%%%%%%%%%%%%%%%%%%%%%%%%%%%%%%%%%%%%%%%%%%%%%%%%%%%%%%%%%%
\clearpage
%%%%%%%%%%%%%%%%%%%%%%%%%%%%%%%%%%%%%%%%%%%%%%%%%%%%%%%%%%%
%%%%%%%%%%%%%%%%%%%%%%%%%%%%%%%%%%%%%%%%%%%%%%%%%%%%%%%%%%%
\clearpage
\begin{deluxetable}{cccc}
\tablewidth{0pt}
\setlength{\tabcolsep}{5mm}
\tablecaption{Filtering scale, ensemble averages of the angles between the satellite positions around the isolated galactic 
hosts and the major, intermediate and minor principal axes of the local velocity shear tensors, respectively.}
\tablehead{$R_{f}$ & $\langle\phi_{1}\rangle$ &  $\langle\phi_{2}\rangle$ & $\langle\phi_{3}\rangle$ \\
($h^{-1}$Mpc) & (degree) & (degree) & (degree)}
\startdata
$1$  & $45.99 \pm 0.22$ & $44.85\pm 0.22$ & $44.12\pm 0.22$ \\
$5$  & $45.96\pm 0.22$ & $44.83\pm  0.22$ & $44.22\pm 0.22$ \\
$10$  & $45.93\pm 0.22$ & $44.65\pm 0.22$ & $44.41\pm 0.22$ \\
$15$  & $45.80\pm0.22$ & $44.70\pm 0.22$ & $44.44\pm 0.22$ 
\enddata
\label{tab:rf}
\end{deluxetable}
%%%%%%%%%%%%%%%%%%%%%%%%%%%%%%%%%%%%%%%%%%%%%%%%%%%%%%%%%%%%
%%%%%%%%%%%%%%%%%%%%%%%%%%%%%%%%%%%%%%%%%%%%%%%%%%%%%%%%%%%
\clearpage
\begin{deluxetable}{cccc}
\tablewidth{0pt}
\setlength{\tabcolsep}{5mm}
\tablecaption{Criterion used to select the galactic satellites around the isolated hosts and the ensemble averages of the three 
alignment angles.}
\tablehead{{\rm criterion} & $\langle\phi_{1}\rangle$ &  $\langle\phi_{2}\rangle$ & $\langle\phi_{3}\rangle$ \\
 & (degree) & (degree) & (degree)}
\startdata
$r_{s}\le R_{\rm v}$  & $45.27\pm 0.43$ & $45.09\pm 0.43$ & $44.64\pm 0.42$ \\
$r_{s} > R_{\rm v}$  & $47.02\pm 0.71$ & $44.49\pm 0.68$ & $43.57\pm 0.67$ \\
\\
\hline
\\
$M_{r}\le -18.2$  & $46.70\pm 0.52$ & $44.60\pm 0.50$ & $43.96\pm 0.50$ \\
$M_{r}> -18.2$  & $45.19\pm 0.62$ & $45.27\pm 0.62$ & $44.23\pm 0.61$ \\
\\
\hline
\\
spirals  & $46.11 \pm 0.29$ & $44.66\pm 0.29$ & $44.26 \pm 0.29$\\
ellipticals  & $45.65 \pm 1.11$ & $45.37 \pm 1.11$ & $44.00\pm 1.07$\\
\enddata
\label{tab:satellite}
\end{deluxetable}
%%%%%%%%%%%%%%%%%%%%%%%%%%%%%%%%%%%%%%%%%%%%%%%%%%%%%%%%%%%%
%%%%%%%%%%%%%%%%%%%%%%%%%%%%%%%%%%%%%%%%%%%%%%%%%%%%%%%%%%%
\clearpage
\begin{deluxetable}{ccccc}
\tablewidth{0pt}
\setlength{\tabcolsep}{5mm}
\tablecaption{Criterion used to select the isolated galactic systems, number of the selected galactic systems, and the ensemble averages 
of the three alignment angles.}
\tablehead{{\rm criterion} & $N_{h}$ & $\langle\phi_{1}\rangle$ &  $\langle\phi_{2}\rangle$ & $\langle\phi_{3}\rangle$ \\
& &(degree) & (degree) & (degree)}
\startdata
$R_{\rm v}\le 0.285\,h^{-1}$Mpc  & $5296$ & $45.83\pm 0.71$ & $44.56\pm 0.70$ & $44.50\pm 0.70$\\
$R_{\rm v}> 0.285\,h^{-1}$Mpc  & $5350$ & $46.09\pm 0.69$ & $45.10\pm 0.67$ & $43.93\pm 0.66$\\
\\
\hline
\\
$M_{r}\le -20.3$  & $5286$ & $46.41\pm 0.70$ & $44.69\pm 0.68$ & $44.07\pm 0.67$\\
$M_{r}> -20.3$  & $5360$ & $45.52\pm 0.70$ & $44.97\pm 0.69$ & $44.36\pm 0.68$\\
\\
\hline
\\
spirals  & $6383$ & $46.39\pm 0.56$ & $44.75\pm 0.54$ & $43.97\pm 0.53$\\
ellipticals & $4263$ & $45.32\pm 0.91$ & $44.95\pm 0.90$ & $44.59\pm 0.90$\\
\\
\hline
\\
in the sheets  & $4422$ & $46.01\pm 1.02$ & $45.05\pm 1.01$ & $44.23\pm 0.99$\\
in the filaments  & $4975$ & $45.84\pm 0.71$ & $44.59\pm 0.69$ & $44.24\pm 0.69$\\
\enddata
\label{tab:host}
\end{deluxetable}
%%%%%%%%%%%%%%%%%%%%%%%%%%%%%%%%%%%%%%%%%%%%%%%%%%%%%%%%%%%%
\end{document}